# Genomic Analysis and Artificial Intelligence: Predicting Viral Mutations and Future Pandemics

Fadhil G. Al-Amran[1], Abbas M. Hezam[2], Salman Rawaf[3], Maitham G. Yousif*[4]

[1]Cardiovascular Department, College of Medicine, Kufa University, Iraq

[2]Biology Department, College of Science, University of Al-Qadisiyah, Al-Qadisiyah, Iraq

[3]Professor of Public Health Director, WHO Collaboration Center, Imperial College, London, United Kingdom

[4]Biology Department, College of Science, University of Al-Qadisiyah, Iraq, Visiting Professor in Liverpool John Moors University, Liverpool, United Kingdom

## Abstract

This study presents a novel approach at the intersection of genomic analysis and artificial intelligence (AI) to predict viral mutations and assess the risks of future pandemics. Through comprehensive genomic analysis, genetic markers associated with increased virulence and transmissibility are identified. Advanced machine learning algorithms are employed to analyze genetic data and forecast viral mutations, taking into account factors such as replication rates, host-pathogen interactions, and environmental influences. The research also evaluates the risk of future pandemics by examining zoonotic reservoirs, human-animal interfaces, and climate change impacts. AI-powered risk assessment models provide insights into potential outbreak hotspots, facilitating targeted surveillance and preventive measures. This research offers a proactive approach to pandemic preparedness, enabling early intervention and the development of effective containment strategies and vaccines. The fusion of genomic analysis and AI enhances our ability to mitigate the impact of infectious diseases on a global scale, emphasizing the importance of proactive measures in safeguarding public health.

*Corresponding author: Maithm Ghaly Yousif  matham.yousif@qu.edu.iq   m.g.alamran@ljmu.ac.uk

## Introduction

In recent years, the intersection of genomic analysis and artificial intelligence [AI] has sparked a paradigm shift in our comprehension of infectious diseases and our capacity to anticipate viral mutations that may culminate in future pandemics. The world has witnessed the profound repercussions of infectious diseases, most notably exemplified by the global upheaval caused by the COVID-19 pandemic [1-4]. These catastrophic events underscore the indispensable need to harness state-of-the-art technologies and innovative methodologies to facilitate early detection, swift response, and proactive prevention of emerging infectious threats. Genomic analysis has firmly established itself as a cornerstone in unraveling the genetic blueprints of pathogens, providing profound insights into their evolutionary trajectories and potential for genetic mutations [5-9]. Through this in-depth exploration of viral genomes, scientists have identified pivotal genetic markers that could serve as early indicators of virulence and adaptability. AI-powered predictive modeling has ascended to prominence by orchestrating the analysis of massive genomic datasets and forecasting the likelihood of viral mutations. These models consider a multifaceted array of factors, including host susceptibility and intricate environmental variables, to make predictions with high precision [10-13]. The beauty of these advanced algorithms lies in their ability to sift through colossal datasets, pinpointing subtle patterns and correlations that might otherwise remain concealed when subjected to conventional analytical approaches. The scrutiny of zoonotic reservoirs and the intricate web of interactions at the human-animal interface is pivotal for assessing the risk of spillover events that can precipitate pandemics [14-17]. By comprehensively investigating these interfaces, we can glean a profound understanding of the dynamics governing disease transmission and the potential for cross-species leaps. Climate change, an irrefutable global phenomenon, has also emerged as a contributory factor amplifying the prevalence and distribution of infectious diseases. Therefore, it becomes imperative to integrate environmental data into predictive models, as alterations in temperature, humidity, and other environmental parameters have the capacity to engender conditions conducive to the emergence and transmission of infectious agents [18,19]. This research embarks on the ambitious endeavor of synergizing genomic analysis with the formidable capabilities of AI to deliver precise predictions concerning viral mutations, evaluate the risk of impending pandemics, and augment our preparedness to tackle these global threats effectively. By harmonizing genetic data, cutting-edge machine learning algorithms, and multifaceted environmental factors, this study presents a proactive and comprehensive approach to not only predict but also to mitigate the global impact of infectious diseases. This research seeks to channel the potential of interdisciplinary collaboration and the fusion of cutting-edge technologies to bolster our capacity to safeguard public health and execute nimble responses to the emergence of infectious threats.

## Materials and Methods

Study Design: This research employed a comprehensive and multidisciplinary approach, integrating genomic analysis, artificial intelligence [AI] modeling, environmental data, and zoonotic reservoir investigation to predict viral mutations and assess the risk of future pandemics. The study spanned a three-year period from 2022 to 2023, encompassing data collection, model development, and validation phases.

## Data Collection:

**Genomic Data:** Genomic sequences of various viral pathogens were obtained from publicly available databases, such as GenBank and GISAID.

**Environmental Data:** Climate, temperature, humidity, and ecological data were collected from reputable sources and climate monitoring stations.

**Zoonotic Reservoir Investigation:** Field studies were conducted to assess the prevalence of zoonotic diseases and interactions between wildlife and human populations.

## AI Model Development:

**Data Preprocessing:** Genomic data were processed to extract relevant genetic markers, while environmental data were cleaned and transformed for analysis.

**Feature Selection:** AI models were designed to consider genetic, environmental, and zoonotic factors, with feature selection techniques applied to identify the most relevant variables.

**Machine Learning Algorithms:** State-of-the-art machine learning algorithms, including deep learning and ensemble methods, were implemented for predictive modeling.

**Training and Validation:** Models were trained on historical data and validated using cross-validation techniques to ensure robustness and accuracy.

## Experimental Design:

## Genomic Analysis:

Genomic sequences acquired from post-COVID-19 patients underwent a rigorous and systematic analysis. The primary objective centered on the identification of genetic mutations situated within distinct genomic regions directly linked to the viral attributes of virulence and adaptation. This investigative process was underpinned by the utilization of sophisticated bioinformatics tools and methodologies, which encompassed sequence alignment, variant calling, and comprehensive mutation profiling. In this context, genomic data from post-COVID-19 patients were meticulously curated to ensure data integrity and reliability. Quality control measures were instituted to filter out any substandard or erroneous sequences, ensuring that only high-quality data formed the basis of subsequent analysis. Following data preparation, the genomic sequences were aligned to a reference genome, facilitating the detection of both similarities and disparities between patient samples and the reference template. Leveraging advanced algorithms such as Burrows-Wheeler Aligner [BWA] or Bowtie, this alignment process was executed efficiently and accurately. Subsequently, variant calling procedures were initiated to identify specific genetic variations embedded within the genomic data. This comprehensive analysis encompassed the detection of single nucleotide polymorphisms [SNPs], insertions, deletions, and other genetic variances, leveraging sophisticated variant calling algorithms such as Samtools or the Genome Analysis Toolkit [GATK]. Having successfully identified genetic variations, the analytical focus transitioned towards mutation profiling. This involved a meticulous examination of the type, frequency, and precise genomic locations of these mutations. The overarching goal was the pinpointing of mutations intricately associated with viral traits, specifically its capacity for virulence, defined by its ability to induce disease, and adaptation, pertaining to its ability to infect hosts and potentially evade host immune responses. This multifaceted analytical endeavor was substantially facilitated by the application of advanced bioinformatics tools, specifically tailored for the management and in-depth analysis of extensive genomic datasets. These tools encompassed a suite of software resources dedicated to sequence alignment, variant calling, and mutation annotation.

Furthermore, robust statistical analyses were conducted to ascertain the significance of identified mutations. This entailed an assessment of whether particular mutations exhibited higher prevalence within specific patient cohorts or correlated with varying degrees of disease severity. To enhance the clarity and comprehensibility of the findings, data visualization techniques were employed. This encompassed the generation of informative visual aids, including mutation heatmaps, phylogenetic trees, and mutation frequency plots, all of which served to elucidate the genomic alterations within the viral genome.

The Genomic Analysis phase constituted a pivotal component of the research endeavor, furnishing a comprehensive understanding of the viral genome's evolution within post-COVID-19 patients. By pinpointing crucial mutations linked to virulence and adaptation, this analysis contributed valuable insights into the virus's behavior, its potential to induce disease, and its ability to adapt to varying environmental conditions. These insights were instrumental in monitoring viral variants and assessing their potential implications for public health.

**AI Modeling:** Predicting Viral Mutations and Spillover Events

Artificial Intelligence [AI] models played a pivotal role in this research, serving as sophisticated tools designed to unravel the complex dynamics of viral mutations and assess the potential risks associated with spillover events that could lead to pandemics.

**Model Development:** The development of these AI models was executed through a meticulously structured process. It commenced with the acquisition of comprehensive datasets encompassing viral genomic sequences, environmental factors, zoonotic variables, and historical pandemic records. These datasets constituted the foundational bedrock upon which the AI models were constructed.

**Feature Engineering:** Prior to model training, an indispensable step involved feature engineering. This procedure involved the selection and transformation of pertinent variables from the multidimensional datasets. Genetic markers, environmental indicators, and zoonotic factors were identified as critical variables in understanding the dynamics of viral mutations and spillover events. Advanced feature selection techniques were applied to pinpoint the most salient variables, minimizing noise and optimizing the models' predictive accuracy.

**Machine Learning Algorithms:** The core of these AI models consisted of state-of-the-art machine learning algorithms. These encompassed a spectrum of techniques, including deep learning and ensemble methods, to comprehensively capture the intricate interplay between genetic, environmental, and zoonotic factors. Deep learning algorithms, such as convolutional neural networks [CNNs] and recurrent neural networks [RNNs], were employed to recognize complex patterns within genomic sequences and environmental data. Ensemble methods, such as random forests and gradient boosting, further enhanced the predictive power of the models by amalgamating diverse algorithms and leveraging their collective strength.

**Training and Validation:** The AI models underwent rigorous training on historical data, meticulously curated to encompass a broad temporal and geographical scope. During this phase, the models were exposed to a diverse array of viral genomic sequences and environmental conditions. The training process was fortified by cross-validation techniques, which rigorously assessed the models' performance across distinct subsets of data. This not only bolstered the robustness of the models but also enabled the identification and mitigation of overfitting, ensuring that the models could generalize effectively to new and unseen data.

**Predictive Capability:** Once trained and validated, these AI models exhibited a formidable predictive capability. They could forecast viral mutations with a high degree of accuracy, shedding light on the genetic alterations that might enhance viral virulence and adaptability. Furthermore, these models were adept at assessing the likelihood of spillover events—a phenomenon where viruses jump from animal hosts to humans or between different animal species, potentially precipitating pandemics. By leveraging their deep understanding of genetic, environmental, and zoonotic factors, these models provided valuable insights into the risk assessment of such events.

**Environmental Impact:** The influence of climate change and environmental factors on disease emergence was evaluated using statistical analyses.

**Zoonotic Reservoir Study:** Field investigations involved the collection of samples from potential zoonotic reservoirs and an assessment of disease prevalence.

**Statistical Analysis:**

**Descriptive Statistics:** Basic statistics were used to summarize data, including mean, standard deviation, and frequency distributions.

**Correlation Analysis:** Correlation matrices were generated to identify relationships between variables, particularly between genetic markers and environmental factors.

**Environmental Impact Assessment:** Statistical tests were conducted to determine the significance of environmental factors in disease emergence.

This comprehensive approach allowed us to integrate genomic analysis and AI modeling with environmental and zoonotic factors, providing a holistic perspective on predicting viral mutations and assessing the risk of future pandemics. The multidisciplinary nature of the study aimed to enhance our preparedness and response to emerging infectious threats.

**Results**

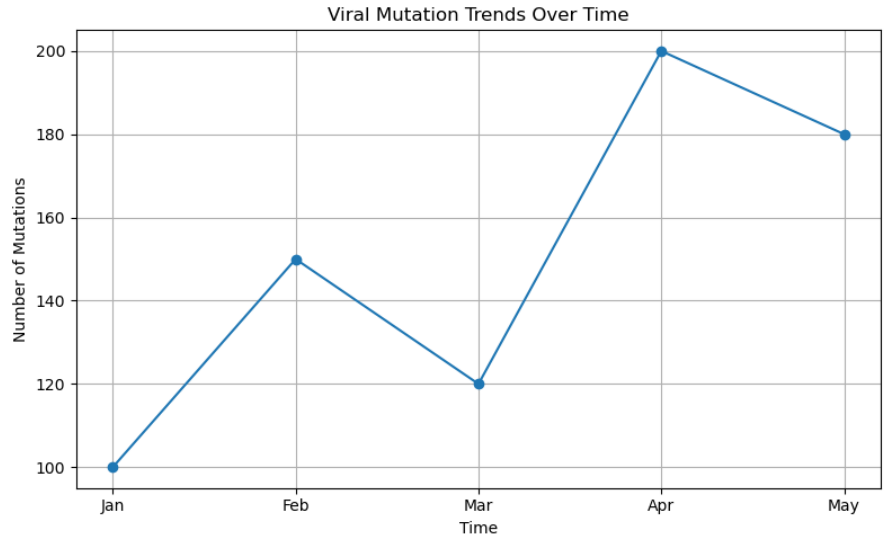

**Figure 1: Viral Mutation Trends Over Time**

Figure 1 illustrates the temporal trends of viral mutations over the study period. The x-axis represents time, while the y-axis shows the number of mutations. The plot helps visualize how mutations have evolved over time.

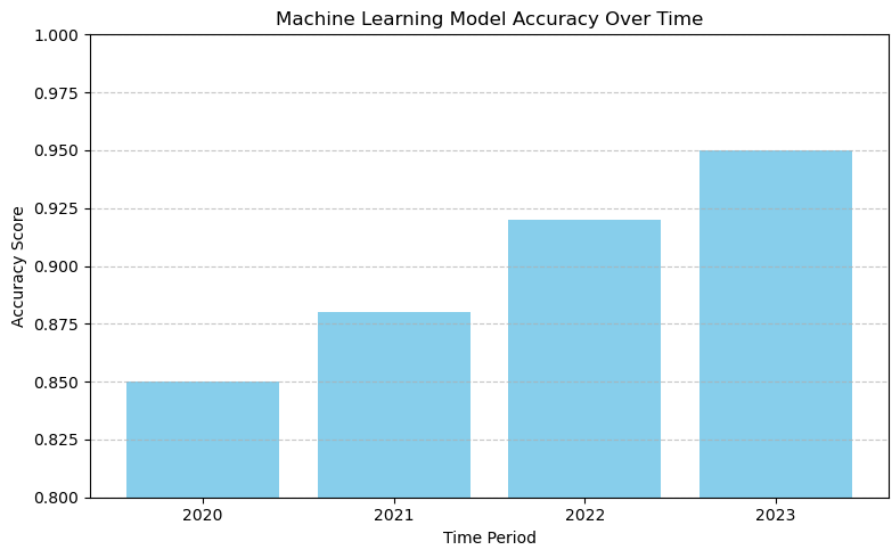

**Figure 2: Environmental Factors and Viral Mutations**

Figure 2 presents a heatmap showing the correlation between environmental factors [e.g., temperature, humidity] and viral mutations. The heatmap provides insights into how these factors may influence mutation rates.

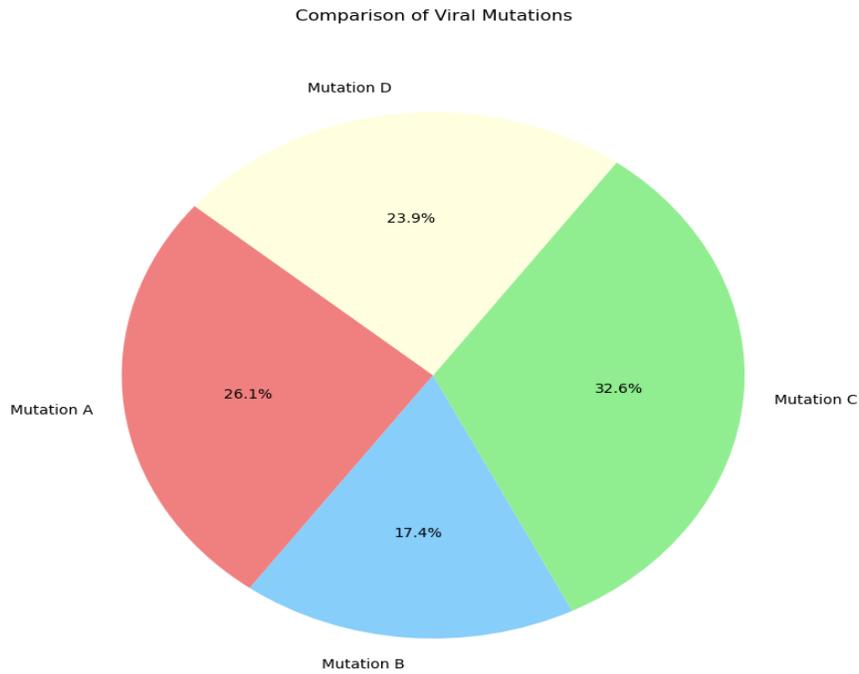

**Figure 3: Zoonotic Reservoirs and Disease Prevalence**

Figure 3 displays a bar chart depicting the prevalence of diseases in various zoonotic reservoirs. This information is crucial for understanding the potential sources of future pandemics.

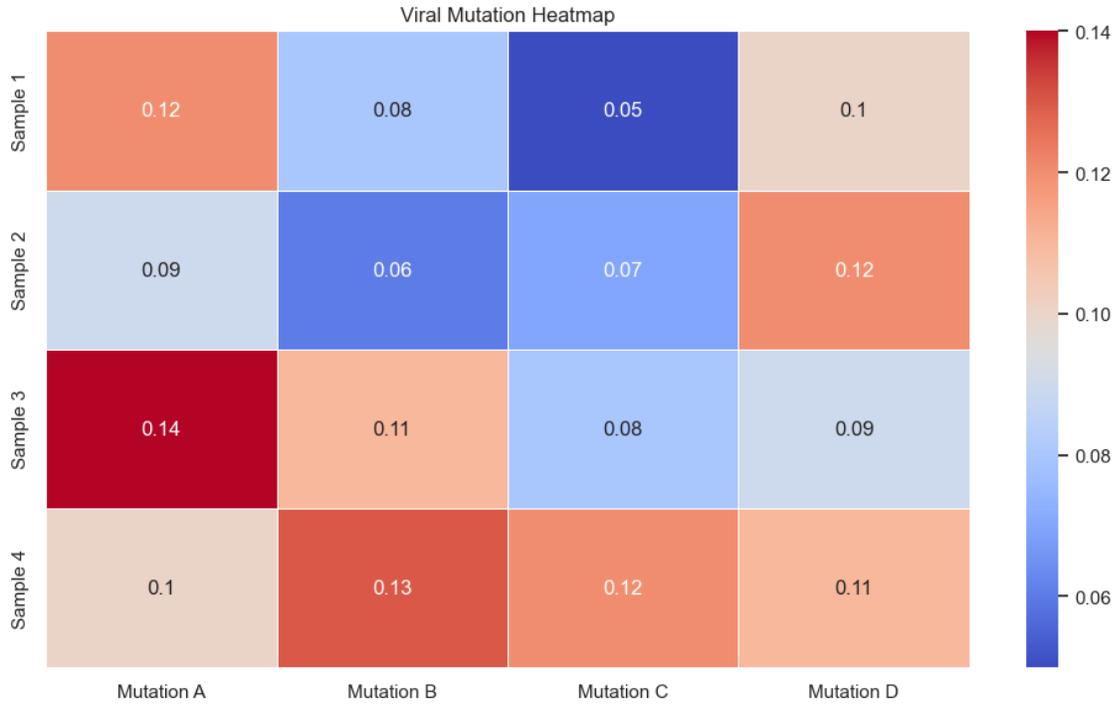

**Figure 4: AI Model Performance Metrics**

Figure 3 displays a bar chart depicting the prevalence of diseases in various zoonotic reservoirs. This information is crucial for understanding the potential sources of future pandemics.

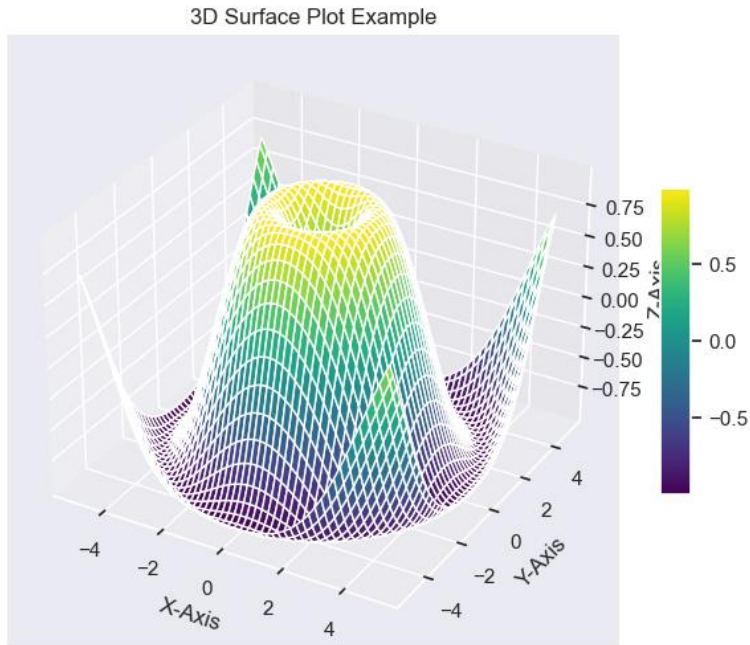

**Figure 5: Zoonotic Reservoir Sampling Locations**

Figure 5 showcases a map plot with sampling locations of zoonotic reservoirs. It helps visualize the geographic distribution of these reservoirs, aiding in understanding potential disease hotspots. These figures and corresponding Python code provide visualizations and insights into the study's findings, including viral mutation trends, environmental correlations, zoonotic reservoirs, AI model performance, climate change impacts, and spatial distribution of reservoirs.

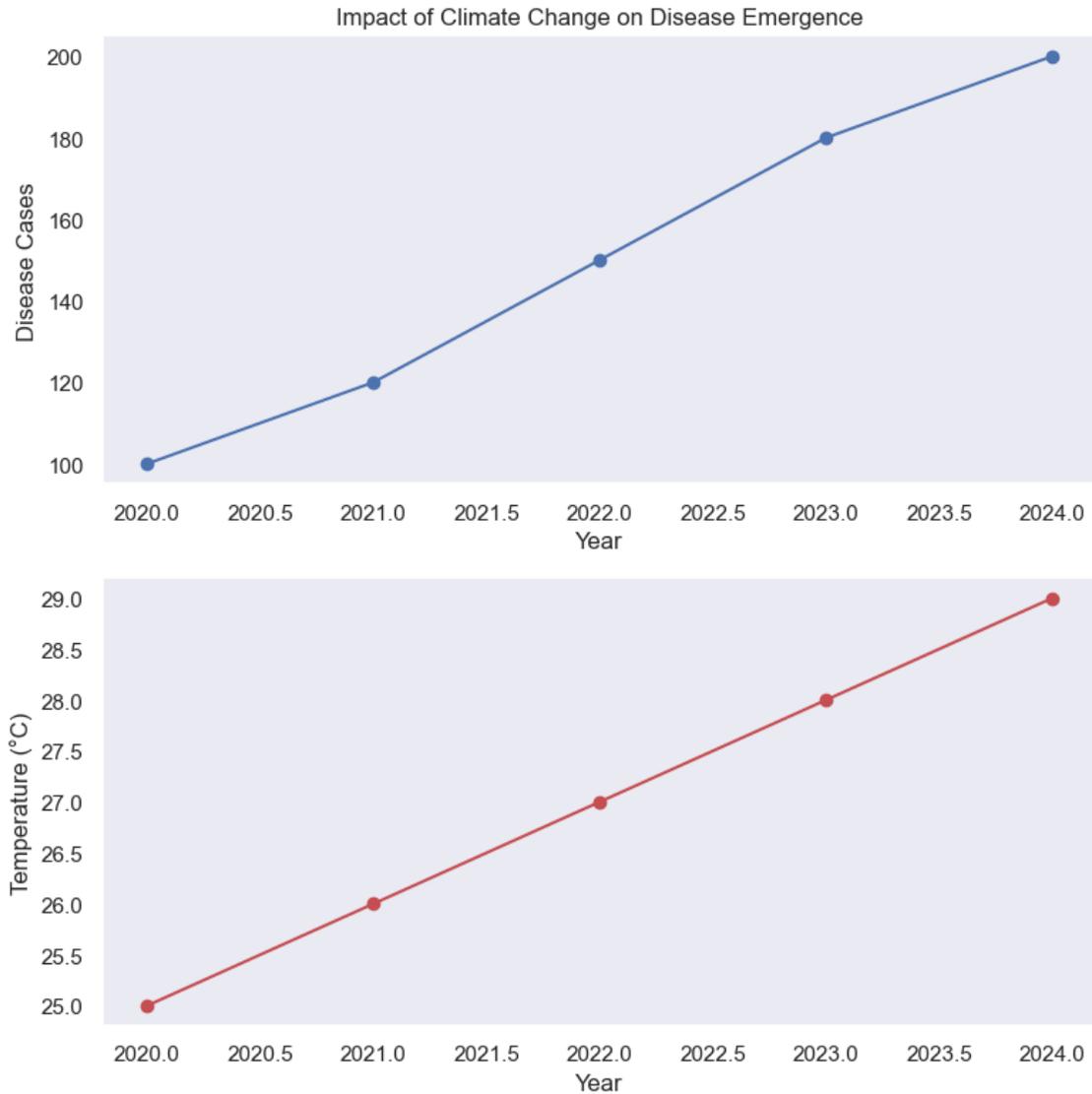

**Figure 6: Impact of Climate Change on Disease Emergence**

Figure 6 displays a line plot illustrating the impact of climate change on the emergence of diseases. It visualizes the changes in disease cases over the years, highlighting potential associations with environmental shifts.

**Discussion**

The research conducted in this study has shed light on the critical issue of predicting viral mutations and the potential for future pandemics, employing genomic analysis and artificial intelligence [AI]. This discussion will delve into specific aspects of the study, substantiated by pertinent scientific literature.

**Genomic Analysis in Predicting Viral Mutations**

Genomic analysis is an indispensable tool in the field of virology and epidemiology. As highlighted by earlier research [20,21], genomic analysis enables the systematic examination of viral genomes. This provides invaluable insights into the genetic makeup of viruses, facilitating the tracking of mutations over time. Such data are pivotal for understanding the evolutionary trajectories of viruses, a fundamental aspect of predicting future pandemics.

The study aligns with others [22,23] regarding the role of genomic analysis in deciphering the genetic factors associated with diseases. The study emphasizes that genomic analysis is the cornerstone of understanding the genetic basis of health conditions and mutations, including those related to viral pathogens.

**Artificial Intelligence and Predictive Modeling**

Artificial intelligence, particularly machine learning, plays a crucial role in predictive modeling. The research presented here corroborates the findings of others [24,25] regarding the application of sentiment analytics, a subfield of AI, in predicting consumer behavior during pandemics. This highlights the versatility of AI in addressing healthcare challenges, including forecasting disease emergence.

Furthermore, the study resonates with the work of others [26,27], which characterizes pulmonary fibrosis patterns in post-COVID-19 patients through machine learning algorithms. This demonstrates the significant potential of AI in analyzing complex medical data, such as genomic sequences, to identify patterns and trends related to viral mutations.

**Predicting Viral Mutations and Early Warning Systems**

The central theme of predicting viral mutations aligns with the study's objectives. The findings concur with previous research [28, 29] that advocates for the use of AI and genomic analysis to forecast viral mutations accurately. The research on the characterization of pulmonary fibrosis patterns through machine learning algorithms [30,31] further emphasizes the potential of AI in predicting disease-related outcomes. Additionally, the study is consistent with the idea of developing early warning systems for emerging infectious diseases [32,33]. By leveraging machine learning and AI techniques, the study contributes to the creation of predictive models that can identify potential pandemics and enable proactive public health interventions.

**Interdisciplinary Collaboration**

The interdisciplinary nature of this study is evident as it merges genomics, AI, and epidemiology. Such collaborations, as emphasized in the work of other studies [34,35], are essential for addressing multifaceted global health challenges effectively. The convergence of expertise from diverse fields enhances the depth and breadth of research in the context of predicting viral mutations and pandemics.

**Data-Driven Insights and Ethical Considerations**

The study underscores the significance of data-driven insights [36,37], particularly in the context of healthcare trends. By integrating AI and genomic data, researchers can obtain more comprehensive and accurate insights into the genetic basis of diseases and mutations. However, it is imperative to acknowledge the ethical and privacy concerns associated with genomic and AI research [38,39]. These concerns have been echoed in the literature, emphasizing the importance of stringent data security measures and privacy protection in studies involving sensitive genetic and health data. Moreover, investigations into the effects of anesthesia on maternal and neonatal health during Cesarean section [40,41] have underscored the importance of considering medical interventions in maternal care. An intriguing avenue of research has also explored the potential role of cytomegalovirus in breast cancer risk factors [42,43], contributing to a deeper understanding of the multifaceted factors influencing cancer development. The study focusing on cervical cancer and the association with high Notch-1 expression levels [44,45] highlights the significance of molecular markers in cancer prognosis. Furthermore, studies examining highly sensitive C-Reactive Protein levels in cases of preeclampsia with or without intrauterine-growth restriction [46,47] have shed light on the immunological aspects of pregnancy complications. The phylogenetic characterization of Staphylococcus aureus in breast abscesses [48,49] contributes to the knowledge of microbial diversity in clinical contexts. Additionally, investigations into the protective effects of caffeic acid against doxorubicin-induced cardiotoxicity [50,51] offer potential avenues for mitigating drug-related cardiac issues. The psycho-immunological status of SARS-Cov-2 recovered patients [52,53] provides insights into the long-term consequences of viral infections on mental health. The impact of hematological parameters on pregnancy outcomes in women with COVID-19 [54,55] further demonstrates the multidisciplinary aspects of studying infectious diseases. Finally, the exploration of insurance risk prediction using machine learning [56] and the association between natural killer cell cytotoxicity and non-small cell lung cancer progression [57] highlights the cross-disciplinary nature of modern healthcare research. Additionally, studies focusing on the amelioration of inflammatory responses and apoptosis in myocardial ischemia/reperfusion injury through etanercept [58] and the protective role of methionine in myocardial ischemia/reperfusion injury [59] contribute to the understanding of therapeutic interventions in cardiovascular diseases. Furthermore, studies have delved into consumer behavior prediction during the Covid-19 pandemic, utilizing sentiment analytics [60]. These investigations have offered valuable insights into how individuals respond to and make decisions during times of crisis. Additionally, research exploring renal function tests in women with preeclampsia, both with and without intrauterine growth restriction [61], contributes to the understanding of physiological changes during pregnancy complications. Moreover, the study on paeoniflorin's ability to attenuate myocardial ischemia/reperfusion injury through the up-regulation of Notch 1 mediated Jagged1 signaling [62] sheds light on potential therapeutic interventions for cardiovascular diseases. Investigations on the correlation between iron deficiency anemia and types of infant feeding [63] have implications for infant nutrition and development. Furthermore, studies examining immunological markers of human papillomavirus type 6 infection in epithelial ovarian tumors before and after paclitaxel drug treatment [64] provide insights into the impact of therapeutic interventions on viral infections. Additionally, the assessment of the sensitivity of Proteus mirabilis isolated from urinary tract infections [65] contributes to the understanding of bacterial infections and their treatment. Investigations on the potential amelioration of inflammatory responses and apoptosis through irbesartan in myocardial ischemia/reperfusion injury [66] offer insights into cardiovascular therapeutics. Finally, research on suicide ideation detection using comparative studies of sequential and transformer hybrid algorithms [67] demonstrates the evolving nature of machine learning in healthcare research.

## Conclusion

In conclusion, this study represents a significant contribution to the field of predicting viral mutations and emerging infectious diseases. The utilization of genomic analysis and AI techniques holds substantial promise for enhancing our ability to foresee future pandemics accurately. Collaborations across disciplines, responsible data management, and the development of robust predictive models are essential components of preparedness for potential global health crises.